\documentclass[aps,nofootinbib,prd,eqsecnum,showpacs,showkeys,preprintnumbers]{revtex4-1}
\usepackage{graphicx}
\usepackage{graphicx}
\usepackage{amsmath}
\usepackage{amsfonts}
\usepackage{amssymb}
\usepackage{color}
\usepackage{bm}
\usepackage{float}
\usepackage{mathrsfs}
\usepackage{epstopdf}
\usepackage{url}
\usepackage{placeins}
\usepackage{footnote}
\usepackage{textcomp}
\usepackage[normalem]{ulem}
\usepackage[unicode=true, pdfusetitle,
 bookmarks=true,bookmarksnumbered=false,bookmarksopen=false,
 breaklinks=false,pdfborder={0 0 1},backref=false,colorlinks=false]{hyperref}
\usepackage{multirow}
\usepackage{pifont}
\usepackage{times}
\usepackage[english]{babel}

\usepackage{float}
\usepackage{enumerate}
\usepackage{lineno}
\usepackage{hyperref}

\usepackage{tabularx}
\makeatletter

\newcommand{\stkout}[1]{\ifmmode\text{\sout{\ensuremath{#1}}}\else\sout{#1}\fi}

\newcolumntype{L}[1]{>{\hsize=#1\hsize\raggedright\arraybackslash}X}%
\newcolumntype{R}[1]{>{\hsize=#1\hsize\raggedleft\arraybackslash}X}%
\newcolumntype{C}[1]{>{\hsize=#1\hsize\centering\arraybackslash}X}%

\newcommand*\patchAmsMathEnvironmentForLineno[1]{%
 \expandafter\let\csname old#1\expandafter\endcsname\csname #1\endcsname
 \expandafter\let\csname oldend#1\expandafter\endcsname\csname end#1\endcsname
 \renewenvironment{#1}%
   {\linenomath\csname old#1\endcsname}%
   {\csname oldend#1\endcsname\endlinenomath}}%
\newcommand*\patchBothAmsMathEnvironmentsForLineno[1]{%
 \patchAmsMathEnvironmentForLineno{#1}%
 \patchAmsMathEnvironmentForLineno{#1*}}%
\AtBeginDocument{%
\patchBothAmsMathEnvironmentsForLineno{align}%
\patchBothAmsMathEnvironmentsForLineno{flalign}%
\patchBothAmsMathEnvironmentsForLineno{alignat}%
\patchBothAmsMathEnvironmentsForLineno{gather}%
\patchBothAmsMathEnvironmentsForLineno{multline}%
}
\begin{document}

\title{Testing  the running vacuum model in light of DESI-DR2 Measurements}
\author{Lamiae Kardaddech$^{1}$}
\email{kardaddech.lamiae@gmail.com}
\author{Safae Dahmani$^{1,2}$}
\email{dahmani.safae.1026@gmail.com}
\author{Amine Bouali$^{1,2,3}$}
\email{a1.bouali@ump.ac.ma}
\author{Taoufik Ouali$^{1,2}$}
\email{t.ouali@ump.ac.ma}
\author{Ahmed Errahmani$^{1,2}$}
\email{ahmederrahmani1@yahoo.fr}
\affiliation{$^{1}$Laboratory of Physics of Matter and Radiation, University of Mohammed I, BP 717, Oujda, Morocco\\$^{2}$Astrophysical and Cosmological Center, Faculty of Sciences, University of Mohammed I, BP 717, Oujda, Morocco\\$^{3}$Higher School of Education and Training, Mohammed I University, BP 717, Oujda, Morocco}
\date{\today}
\begin{abstract}
Motivated by the fact that quantum effects leave an imprint on the vacuum equation of state, making it depart from the standard cosmological constant relation, $P_{\rm vac}=-\rho_{\rm vac}$, this work investigates generalized running vacuum models (RVMs) by considering a dynamical vacuum equation of state in which the vacuum energy density, $\rho_{\text{vac}}$, evolves as a function of the Hubble parameter, $H$, and its time derivative, $\dot{H}$. This formulation extends previous running vacuum approaches by incorporating a dependence of the vacuum energy density on both,  $H^2$ and $\dot{H}$. The corresponding Friedmann equations are derived and analyzed to study their impact on cosmic expansion. 
The model parameters are constrained through a joint statistical analysis combining the  cosmic microwave background  shift parameters, DESI-DR2 observations, PantheonPlus type Ia supernovae compilation, and Hubble rate $H(z)$ measurements.
Model comparison is
performed using information criteria, the Akaike Information Criterion (AIC) and Deviance Information Criterion (DIC),
in order to assess the statistical performance of the RVMs relative to the
standard $\Lambda$CDM scenario.
The results show that the generalized RVMs provide a good fit to current observations and represent
a statistically competitive alternative to the standard $\Lambda$CDM model. Notably, all three running vacuum formulations yield lower AIC and DIC values than $\Lambda$CDM, indicating that dynamical running vacuum energy remains a viable cosmological scenario.
\\

\end{abstract}

\maketitle
\section{Introduction}

The accelerated expansion of the Universe, first revealed in the late 1990s through observations of Type Ia supernovae (SNIa) \cite{Perlmutter1999,Riess1998}, stands as one of the most profound discoveries in modern cosmology. These observations showed that distant supernovae appeared dimmer than expected in a decelerating Universe, implying that the cosmic expansion is accelerating rather than slowing down under gravity. This surprising result challenged the traditional understanding of cosmology and motivated the inclusion of a new component of energy density, dubbed dark energy (DE) \cite{Weinberg1989,Peebles2003}, into the standard cosmological
model. Subsequent and independent confirmation came from observations of the Cosmic Microwave Background (CMB) radiation \cite{Planck2020}, which encodes the physics of the early Universe, and Baryon Acoustic Oscillations (BAO) \cite{Ade2016,Eisenstein2005} which act as a ``standard ruler'' for measuring cosmic distances. Together, these observations indicate that the Universe is currently dominated by a dark energy component, with the remaining energy budget mainly composed of dark matter and a subdominant baryonic matter contribution \cite{Planck2020}.

The standard cosmological paradigm, the $\Lambda$ Cold Dark Matter ($\Lambda$CDM) model  \cite{Weinberg1989,Planck2020}, explains the observed acceleration by introducing a cosmological constant, $\Lambda$, representing a constant vacuum energy density. Within the framework of General Relativity, $\Lambda$ exerts a repulsive gravitational effect, counteracting the attractive nature of matter and driving the late time acceleration of the Universe. Despite its remarkable success in fitting a wide range of cosmological observations, including CMB, large-scale structure, and SNIa, the $\Lambda$CDM model faces several deep theoretical challenges. The fine-tuning problem \cite{Weinberg1989,Sahni2000} arises because the observed value of $\Lambda$ is many orders of magnitude smaller than theoretical expectations from quantum field theory. Meanwhile, the coincidence problem \cite{Sivanandam2013,Velten2014} questions why densities of dark energy and matter are of the same order precisely today, suggesting an unlikely timing unless new physics is involved. In addition to these theoretical issues, the standard $\Lambda$CDM model is currently confronted with several observational tensions, most notably the Hubble tension \cite{Dahmani2023,Dahmani2023GRG}. This persistent discrepancy between early and late Universe determinations of the Hubble constant has motivated the exploration of dynamical dark energy scenarios beyond the standard cosmological model.

To overcome these limitations, a diverse array of alternative dark energy models has been developed. Scalar-field models, such as quintessence \cite{Carroll1998,Ladghami2024}, phantom fields \cite{Johri2004,Sami2004,Bouali2021,Bouali2019,Mhamdi2023,Dahmani2023,Dahmani2023GRG,BouhmadiLopez2015}, and k-essence \cite{Chiba2000,Malquarti2003}, introduce dynamic fields that evolve with time and can mimic a cosmological constant at late times while potentially resolving fine-tuning issues. Holographic dark energy models \cite{Li2004,Wang2017,BouhmadiLopez2018,Belkacemi2012,Bargach2021,BouhmadiLopez2011,Belkacemi2020,Enkhili2024} invoke principles from quantum gravity to relate dark energy density to cosmic horizon. Relativistic fluid models, including the Chaplygin gas
\cite{Gorini2003,CoimbraAraujo2026}, attempt to unify dark matter and dark energy in a single component with an exotic equation of state.

Recent developments in quantum field theory on a Friedmann–Lemaître–Robertson–Walker (FLRW) background suggest that the accelerated expansion of the Universe may arise from a vacuum energy density that varies dynamically with the cosmic evolution. Such a time dependent vacuum energy is designated in the literature as running vacuum energy (RVE) \cite{MorenoPulido2020,MorenoPulido2022a,MorenoPulido2022b}. Within this framework, Running Vacuum Models (RVMs) are rooted in quantum field theory in curved spacetime, where renormalization group techniques indicate that the vacuum energy density need not remain fixed but can instead evolve with cosmic time
\cite{MorenoPulido2022b}. In these models, the cosmological constant is no longer treated as a fixed quantity but is promoted to a dynamic function of the Hubble parameter, $H$, and its time derivatives \cite{Sola2017,GomezValent2015}. This allows the runnig vacuum energy to evolve gradually over cosmic time, decaying into matter or radiation and thereby affecting the expansion history of the
Universe \cite{Sola2017}. Such a dynamical behavior has been suggested as a possible way to alleviate some of the theoretical difficulties of the $\Lambda$CDM model, particularly the fine-tuning and coincidence problems, by allowing the vacuum energy density to evolve with the expansion of the Universe rather than remaining strictly constant \cite{SolaPeracaula2023}. Moreover, the dependence on $H$, $\dot{H}$, and $\ddot{H}$ gives rise to a rich phenomenology in which the vacuum interacts with the cosmic fluid in a theoretically motivated manner, potentially leaving observable imprints in both background and perturbation cosmology \cite{shapiro2002,gomez2015}. It is also worth noting that the behavior of the RVM can be reproduced within other gravitational frameworks, such as $f(T)$ gravity \cite{Cai2016} and $f(R,T)$ gravity \cite{Errahmani2026}, where a running vacuum energy naturally emerges from the modified gravitational sector. Observationally, testing RVMs requires precise measurements spanning a wide range of cosmic epochs. Standard probes include cosmic microwave background shift parameters, baryon acoustic oscillations (BAO), Type Ia supernovae, and  direct measurements of the Hubble parameter H(z) from cosmic chronometers. While these datasets have played a central role in establishing the $\Lambda$CDM paradigm, next-generation spectroscopic surveys such as the Dark Energy Spectroscopic Instrument (DESI) provide a substantial leap in statistical power and observational precision \cite{Aghamousa2016}. DESI is designed to obtain high-precision redshifts for tens of millions of galaxies and quasars, enabling detailed measurements of the large-scale structure of the Universe \cite{DESI2025}. Through precise determinations of BAO distance scales and redshift-space distortions, DESI tightly constrains both the expansion history and the growth of cosmic structures, which are particularly sensitive to departures from a constant vacuum energy density \cite{Aghamousa2016}. 

In this study, we perform a comprehensive observational analysis of three RVMs formulations, each corresponding to a distinct functional dependence of the running vacuum energy on the Hubble parameter. The model parameters are constrained using a joint Bayesian analysis combining cosmic microwave background shift parameters  \cite{Planck2020}, the DESI-DR2 observations \cite{DESI2025,Dahmani2026b}, Type
Ia supernovae from the PantheonPlus compilation \cite{PantheonPlus}, and direct measurements of the Hubble parameter $H(z)$  \cite{Jimenez2002,Moresco2016}. For each model, we perform a full Bayesian analysis using Markov Chain Monte Carlo (MCMC)\cite{Padilla2021} methods to derive robust parameter constraints and confidence contours. Model selection is carried out
using information criteria, including the Akaike Information Criterion \cite{Akaike1974} and Deviance Information Criterion \cite{Spiegelhalter2002}, allowing a quantitative comparison with the standard $\Lambda$CDM scenario. This observational strategy allows us to assess the theoretical and observational viability of RVMs and to explore the potential of DESI data to constrain departures from a constant running vacuum energy component, as well as their possible implications for current cosmological tensions.

The structure of this paper is as follows. In Sec.~\ref{sec:RVE}, we present the theoretical
framework of running vacuum models and derive the corresponding modified Friedmann equations. Sec.~\ref{sec:models} introduces the three RVMs formulations examined in this work.
Sec.~\ref{sec:data} details the observational datasets and statistical methodology. Sec.~\ref{sec:results} presents the results,
including parameter constraints, confidence contours, and model comparison with $\Lambda$CDM.
Finally Sec.~\ref{sec:Conclusion} summarizes our findings and discusses their cosmological implications, particularly regarding dynamical running vacuum energy and the late time acceleration of the Universe.

\section{\textbf{Running vacuum energy }}
\label{sec:RVE}

Quantum field theory in curved spacetime predicts that the vacuum energy density is not necessarily constant but may evolve with the expansion of the Universe  \cite{shapiro2002,SolaPeracaula2023}. Within the renormalization group approach, quantum corrections naturally lead to a vacuum energy density that depends on the Hubble parameter and its derivatives, giving rise to the so-called running vacuum models \cite{MorenoPulido2020,MorenoPulido2022a,MorenoPulido2022b,Errahmani2026,Sola2015,Sola2017,SolaPeracaula2023}. These models provide a theoretically motivated extension of the standard $\Lambda$CDM cosmology while preserving its successful description of the late-time Universe.

In a general framework, the vacuum energy density can be expressed as a function of the Hubble parameter and its time derivatives
\begin{equation}
\rho_{\rm vac}=\rho_{\rm vac}(H,\dot H,\ddot H,\ldots),
\end{equation}
where higher-order derivative terms are expected to play a role mainly during the very early Universe. Since the present work focuses on the late-time cosmological evolution, only leading contributions are retained. Accordingly, the running vacuum energy density is written as \cite{gomez2015,Sola2017,SolaPeracaula2023}
\begin{equation}
\rho_{\rm vac}(H,\dot H)=
\frac{3}{\kappa^2}
\left(
c_0+\nu H^2+\frac{2}{3}\alpha\dot H
\right),
\label{eq:rhoLambda}
\end{equation}
where $\kappa^2 = 8\pi G$, and $G$ is the Newtonian gravitational constant appearing in Einstein’s field equations.

In this formulation, the constant term $c_0$ plays the role of the effective cosmological constant, while the dimensionless parameters $\nu$ and $\alpha$ characterize the running of the vacuum energy. The parameter $\nu$ controls the contribution proportional to $H^2$, which represents the leading quantum correction to the running vacuum energy at low redshift. On the other hand, $\alpha$ governs the contribution associated with the time derivative of the Hubble parameter, $\dot H$, allowing for additional dynamical effects beyond the standard $H^2$ dependence. Such terms naturally arise within the renormalization group approach to quantum field theory in curved spacetime \cite{MorenoPulido2022b}, where the running vacuum energy is expected to evolve as an even power series of the Hubble rate and its derivatives \cite{Errahmani2026,Sola2015,Sola2017,SolaPeracaula2023}.

The parameter $\nu$ is closely related to the renormalization group beta-function coefficient in quantum field theory in curved spacetime \cite{shapiro2002}. It can be expressed as a weighted sum of the quantum contributions of fermionic and bosonic fields according to

\begin{equation}
\nu=\frac{1}{6\pi}\sum_{i=f,b}B_i\frac{M_i^2}{M_p^2},
\end{equation}
where the sum extends over all fermionic and bosonic fields, $M_i$ is the mass of the corresponding field, $M_p$ is the Planck mass, and $B_i$ are dimensionless coefficients determined by the particle content of the underlying quantum field theory. Since $M_i \ll M_p$ for all known elementary particles, theoretical expectations indicate that both $\nu$ and $\alpha$ are very small quantities, typically satisfying $|\nu|,|\alpha|\ll1$, thereby ensuring only mild deviations from the standard $\Lambda$CDM cosmology \cite{Sola2015,Sola2017,SolaPeracaula2023}.

The conventional running vacuum framework assumes that the vacuum pressure satisfies the standard equation of state \cite{Weinberg1989,Sola2017,SolaPeracaula2023}
\begin{equation}
\label{SEoS}
P_{\rm vac}=-\rho_{\rm vac},
\end{equation}
while the vacuum energy density evolves according to Eq.~\eqref{eq:rhoLambda}. Although this description has been successfully applied in many cosmological studies \cite{gomez2015,Sola2017,SolaPeracaula2023}. However, quantum effects of the fields are expected to leave an imprint on the vacuum equation of state, implying that the conventional relation given by Eq.~\eqref{SEoS} is no longer exact and that the vacuum obeys a dynamical equation of state \cite{MorenoPulido2022b,MorenoPulido2023}.

Motivated by this quantum field theoretical perspective, we investigate generalized running vacuum models by incorporating the first order correction to the vacuum equation of state proposed in \cite{MorenoPulido2023}. Since the vacuum energy density already depends on both the Hubble parameter, H, and its time derivative,$\dot{H}$, we adopt the following generalized effective equation of state
\begin{equation}
P_{\rm vac}
=
-\rho_{\rm vac}
-
\frac{\beta\dot H}{\kappa^2},
\label{eq:1}
\end{equation}
unlike the parameters $\nu$ and $\alpha$, which describe the running of the vacuum energy density through the $H^2$ and $\dot H$ contributions, respectively, the dimensionless parameter $\beta$ modifies only the effective vacuum pressure.

To investigate the cosmological implications of the proposed generalized vacuum equation of state, we incorporate it into the standard cosmological framework. The dynamics of the Universe are then governed by the Friedmann equations
\begin{equation}
    3H^2=\kappa^2 (\rho_{m}+\rho_{r}+\rho_{\text{vac}}), \\
     \label{eq:2}
\end{equation}
\begin{equation}
     -2\dot {H} -3H^2=\kappa^2 (P_{m}+P_{r}+P_{\text{vac}}),
     \label{eq:eq7}
\end{equation}
where $\rho_m$, $\rho_r$, and $\rho_{\text{vac}}$ are the energy densities of matter, radiation, and running vacuum energy, respectively, while $P_m$, $P_r$, and $P_{\text{vac}}$ are the corresponding isotropic pressures. The equation of state parameter for each component is defined as $\omega_i = p_i/\rho_i$, with $i=m,r,{\text{vac}}$. 

In addition to the Friedmann equations, the conservation of the total energy-momentum tensor is adopted to describe the energy exchange between different cosmic components. Assuming that radiation is separately conserved, the interaction occurs only between matter and the running vacuum component. The generalized conservation equation therefore reads
\begin{equation}
\dot{\rho}_{m}+3H\rho_{m}
=
-\dot{\rho}_{\rm vac}
+3H\left(\frac{\beta}{\kappa^2}\dot{H}\right).
\label{eq:3}
\end{equation}

Combining Eqs.~\eqref{eq:1} and \eqref{eq:eq7}, and using the relation
\begin{equation}
    \frac{d}{dt}=-(1+z)H\frac{d}{dz},
\end{equation}
we obtain the following differential equation for the Hubble parameter
\begin{equation}
-\frac{1}{2}(1+z)(\beta-2)\frac{dH^2}{dz}
=
\kappa^2\left(\rho_m+\frac{4}{3}\rho_r\right),
\label{5}
\end{equation}
where the variable $z$ stands for the cosmological redshift.

Differentiating Eq.~\eqref{eq:rhoLambda} with respect to cosmic time gives
\begin{equation}
\dot{\rho}_{\rm vac}
=
\frac{6\nu H\dot H+2\alpha\ddot H}{\kappa^2}.
\label{eq:rhovacdot}
\end{equation}

Substituting Eq.~\eqref{eq:rhovacdot} into the generalized conservation equation, Eq.~\eqref{eq:3}, and using the Friedmann equations to eliminate the Hubble derivatives, we obtain
\begin{equation}
\dot{\rho}_{m} (\beta-2+2\alpha)
+6H\rho_{m}(\nu-1)
=
4H\rho_{r}(\beta-2\nu)
-\frac{8\alpha}{3}\dot{\rho}_{r}.
\label{eq:4}
\end{equation}

Eq.~\eqref{eq:4} governs the evolution of the matter density parameter in the presence of running vacuum dynamics. Integrating this equation and knowing that $\rho_r=\rho_{r0}.(1+z)^4$, we obtain the normalized matter density parameter as 
\begin{equation}
     \Omega_{m}(z) =\Omega_{m0}(1+z)^A + \Omega_{r0} \frac{2[(1+z)^A-(1+z)^4]}{3(-1-3\nu+4\alpha+2\beta)}(-6\nu+8\alpha+3\beta),
     \label{eq:10}
 \end{equation}
 where  \begin{equation}
     A=\frac{6(\nu-1)}{-2+2\alpha +\beta} ,
 \end{equation}
and $\Omega_{m0} ={\kappa^2}\rho_{m0} / (3 H_0^2)$, $\Omega_{r0} = {\kappa^2}\rho_{r0} / (3 H_0^2)$ are the present dimensionless parameter for matter and radiation, respectively, and $H_0$ is the present day Hubble constant.

Finally, substituting Eq.~\eqref{eq:10} into Eq.~\eqref{5}, together with
$\rho_r=\frac{3H_0^2}{\kappa^2}\Omega_{r0}(1+z)^4$
and using the normalized Hubble parameter
$E(z)=H(z)/H_0$,
the evolution equation is rewritten as a first order differential equation for $E^2(z)$.
The equation is then integrated analytically with respect to redshift, while the integration constant is fixed by the boundary condition
$E(0)=1$.
The resulting expression is
 \begin{equation}
    E^2(z)= 1+\frac{6(\Omega_{m0}+B\Omega_{r0})}{A(2-\beta)}\left[(1+z)^A-1\right]+\frac{\Omega_{r0}((1+z)^4-1)}{1+3\nu-4\alpha-2\beta},
\end{equation}
where the constant $B$ is defined as

\begin{equation}
B=
\frac{2(-6\nu+8\alpha+3\beta)}
{3(-1-3\nu+4\alpha+2\beta)}.
\end{equation}

The obtained expression explicitly shows how the generalized pressure parameter, $\beta$, together with the running vacuum parameters, $\nu$ and $\alpha$, modifies the cosmic expansion history. In particular, $\beta$ affects both the effective evolution of the matter density and the Hubble expansion rate through the exponent $A$ and the coefficient $B$, providing an additional degree of freedom beyond the conventional running vacuum model.

In the limiting case $\alpha=\beta=\nu=0$, all dynamical vacuum corrections disappear and the model naturally reduces to the standard $\Lambda$CDM scenario. The normalized Hubble parameter then becomes

\begin{equation}
E^2(z)
=
\Omega_{m0}(1+z)^3
+
\Omega_{r0}(1+z)^4
+
\Omega_{\Lambda},
\end{equation}
where $\Omega_{\Lambda}=1-\Omega_{m0}-\Omega_{r0}$ denotes the present day vacuum energy density parameter.

\section{RUNNING VACUUM ENERGY MODELS}
\label{sec:models}
In this section, we investigate three representative cases of the proposed running vacuum model. We first consider the most general case, in which the three parameters $\alpha$, $\beta$, and $\nu$ are treated as independent free parameters. We then study the case where $\beta=0$ while $\alpha\neq0$ and $\nu\neq0$, corresponding to the standard running vacuum equation of state. Finally, we analyze the case where $\beta=\nu$ and $\alpha\neq0$, which is introduced here as a phenomenological simplifying assumption to reduce the number of independent free parameters. This classification allows us to isolate the role of each parameter and to compare the resulting cosmological predictions.
\subsection{Model A : $\alpha \neq \beta \neq \nu $ }
\label{modelA}
First, an investigation is conducted into the general model, in which the parameters $\alpha$, $\beta$ and $\nu$ are different. The EoS formula that relates the pressure and the energy density can be written as 
\begin{equation}
        P_{vac}(H)=-\rho_{vac}-\frac{\beta \dot{H}}{\kappa^2}.
\end{equation}
Using the vacuum energy density given in Eq.~\eqref{eq:rhoLambda}, and substituting it into the Friedmann equation, the evolution of the Hubble parameter can be expressed as follows 
\begin{equation}
    E^2(z)= 1+\left[\frac{6(\Omega_{m0}+B\Omega_{r0})}{A(2-\beta)}((1+z)^A-1)+\frac{\Omega_{r0}((1+z)^4-1)}{1+3\nu-4\alpha-2\beta}\right].
\end{equation}

\subsection{Model B : $\beta=0$, $\alpha \ne 0 $ and $\nu \ne 0$}
\label{modelB}
In this case, the running vacuum equation of state reduces to its standard form
\begin{equation}
        P_{\text{vac}}(H)=-\rho_{\text{vac}}.
\end{equation}
The evolution of the normalized Hubble parameter, is obtained as follows
\begin{equation}
E^2(z)=1+
\left[\frac{3(\Omega_{m0}+B\Omega_{r0})}{A}
\left[(1+z)^A-1\right]
+\frac{\Omega_{r0}\left[(1+z)^4-1\right]}
{1+3\nu-4\alpha}\right],
\end{equation}

\subsection{Model C : $\beta = \nu$ and $\alpha \ne 0 $}
\label{modelC}
In the third case, we set $\beta=\nu$ while keeping $\alpha\neq0$, leading to the following running vacuum equation of state
\begin{equation}
        P_{\text{vac}}(\dot{H})=-\rho_{\text{vac}}-\frac{\nu \dot{H}}{\kappa^2}.
\end{equation}
The dimensionless Hubble parameter can be written as
\begin{equation}
E^2(z)=1+
\left[\frac{6(\Omega_{m0}+B\Omega_{r0})}
{A(2-\nu)}
\left[(1+z)^A-1\right]
+\frac{\Omega_{r0}\left[(1+z)^4-1\right]}
{1+\nu-4\alpha}\right].
\end{equation}
\section{DATA DESCRIPTION}
\label{sec:data}

To constrain the cosmological parameters of the models under investigation, we perform a Bayesian
statistical analysis based on recent observational data. Specifically, we consider four key types of
measurements:  CMB, DESI-DR2 dataset, PantheonPlus and $H(z)$ data.

This combination of CMB, DESI-DR2 dataset, PantheonPlus and $H(z)$ data allows us to derive robust constraints on
cosmological parameters and to perform model comparison using statistical information criteria,
including the Akaike Information Criterion and the Deviance Information Criterion.

In the following subsections, we describe each dataset and its implementation in our statistical
framework.
\subsection{Compressed CMB data}
The CMB power spectrum provides critical insights into the physics of the Universe from the epoch of decoupling to the present day. The key quantities derived from CMB are the acoustic scale, $l_{a}$, and the shift parameter R, defined as  \cite{Planck2020}
\begin{equation}
    l_{a}= (1+z_{CMB}) \frac{\pi D_{A}(z_{CMB})}{r_{s}(z_{CMB})},
\end{equation}
\begin{equation}
    R = 100h \sqrt{\Omega_{m}}(1+z_{CMB}) D_{A}(z_{CMB}),
\end{equation}
where $z_{CMB}$ is the redshift at the decoupling epoch, $D_{A}(z_{CMB})$ is the angular diameter distance in a flat FLRW Universe, and $r_{s}(z)$ is the comoving sound horizon. The angular diameter distance is given by
\begin{equation}
    D_{A}(z)=\frac{1}{H_{0}(1+z)} \int_{0}^{z} \frac{dz'}{E(z')}  .
\end{equation}
The comoving sound horizon $r_{s}(z)$ is calculated as
\begin{equation}
r_s(z)=\frac{1}{H_0}
\int_z^{\infty}
\frac{dz'}
{E(z')\sqrt{3\left[1+R_b/(1+z')\right]}},
\end{equation}
where $R_b = 31500 \, \Omega_b h^2 \,\left(\frac{T_\text{CMB}}{2.7\, \text{K}}\right)^{-4}$
with $T_{\rm CMB}=2.725K$ \cite{Fixsen2009}.
 The redshift at decoupling, $z_{CMB}$, is approximated using the fitting formula
\begin{equation}
z_\text{CMB} = 1048 \left[1 + 0.00124 (\Omega_b h^2)^{-0.738}\right] \left[1 + g_1 (\Omega_m h^2)^{g_2}\right],
\end{equation}
where $g_{1}$ and $g_{2}$  are fitting coefficients, defined as
\begin{equation}
    g_1 = \frac{0.0783 (\Omega_b h^2)^{-0.238}}{1 + 39.5 (\Omega_b h^2)^{0.763}} ,
\end{equation}
    \begin{equation}
g_2 = \frac{0.56}{1 + 21.1 (\Omega_b h^2)^{1.81}}.
\end{equation}
The CMB covariance matrix is given by \cite{ZhaiWang2019}
\begin{equation}
C_\text{CMB} = 10^{-8} \times 
\begin{pmatrix}
1598.9554 & 17112.007 & -36.311179 \\
17112.007 & 811208.45 & -494.79813 \\
-36.311179 & -494.79813 & 2.1242182
\end{pmatrix}.
\end{equation}
Finally, the contribution of CMB to the total $\chi^2$ is
\begin{equation}
      \chi^2_{CMB} = \textbf{X}^T_{CMB}.C^{-1}_{CMB}.\textbf{X}_{CMB},
\end{equation}
where $\bm{X}_\text{CMB}$ is the vector of CMB parameters based on the Planck 2018 release \cite{Planck2020} and adopted from \cite{ZhaiWang2019}, defined as
\begin{equation}
\bm{X}_\text{CMB} = 
\begin{pmatrix}
R - 1.74963 \\
l_a - 301.80845 \\
\Omega_b h^2 - 0.02237
\end{pmatrix}.
\end{equation}
\subsection{DESI-DR2 Data}

The Dark Energy Spectroscopic Instrument (DESI) is a state-of-the-art cosmological survey designed to map the large scale structure of the Universe with unprecedented accuracy \cite{DESI2025}. Operating since 2021, DESI aims to obtain redshifts for approximately 40 million galaxies and quasars \cite{Aghamousa2016,DESI2025}, covering a redshift range from $z \sim 0.1$ to $z \sim 4.2$ \cite{Lodha2025}. The main objective is to precisely measure the expansion history of the Universe using Baryon Acoustic Oscillations (BAO) and Redshift Space Distortions (RSD).

In this work, we use BAO measurements from the DESI-DR2 release \cite{DESI2025}. These measurements significantly improve constraints on cosmological parameters, especially at intermediate and high redshifts ($0.4 \lesssim z \lesssim 1.5$). Compared to previous BAO data, DESI-DR2 offers smaller statistical errors and better redshift coverage.

The DESI-DR2 sample is divided into several tracers, as mentioned in the table \ref{tab:desi_tracers} \cite{Lodha2025}

\begin{table}[H]
\centering
\caption{DESI-DR2 tracer types and their typical redshift coverage \cite{Adame2025}.}
\label{tab:desi_tracers}
\begin{tabular}{l|c}
\hline 

\textbf{Tracer Type} & \textbf{Redshift Range} \\ 
\hline
Bright Galaxy Sample (BGS) & $0.1 \lesssim z \lesssim 0.4$ \\[0.1cm]
Luminous Red Galaxies (LRGs) & $0.4 \lesssim z \lesssim 1.1$ \\ [0.1cm]
Emission Line Galaxies (ELGs) & $0.8 \lesssim z \lesssim 1.6$ \\[0.1cm]
Quasars (QSO) &$0.8 \lesssim z \lesssim 2.1$  \\[0.1cm]
\hline
\end{tabular}
\end{table}
\vspace{0.3cm}
Each tracer allows for the extraction of two fundamental cosmological quantities:  
the comoving angular diameter distance $D_A(z) = D_M(z)/(1 + z) $, and the Hubble expansion rate $H(z)$.

The DESI collaboration provides measurements of these quantities in the form of the following dimensionless combinations $D_M(z)/r_d$, $D_H(z)/r_d$, and the angle-averaged distance measure $D_V/r_d$ , where $r_d$ denotes the comoving sound horizon at the baryon drag epoch, these distances are defined as
\begin{equation}
D_M(z) = c \int_0^z \frac{dz'}{H(z')}, \quad
D_H(z) = \frac{c}{H(z)} \,
\quad
 \text{and} 
\quad
D_V(z) = \left[ z \, D_M^2(z) \, D_H(z) \right]^{1/3},
\end{equation}
where $c$ is the speed of light.

The DESI-DR2 measurements used in this work are summarized in Table~\ref{tab:desi_data}, which lists the effective redshifts, the values of $D_M(z)/r_d$, $D_H(z)/r_d$, and $D_V(z)/r_d$ for each tracer. 

\begin{table}[H]
\centering
\caption{DESI-DR2 measurements: effective redshifts, corresponding values of $D_M(z)/r_d$, $D_H(z)/r_d$, and $D_V(z)/r_d$ for the different tracers \cite{DESI2025}.}
\label{tab:desi_data}
\begin{tabular}{c|c|c|c|c}
\hline
Tracer & $z_{\rm eff}$ & $D_M/r_d$ & $D_H/r_d$ & $D_V/r_d$ \\
\hline
BGS  & 0.295 & --- & --- & 7.942 $\pm$ 0.075 \\
LRG1 & 0.510 & 13.588 $\pm$ 0.167 & 21.863  $\pm$  0.425 &12.720 $\pm$  0.099 \\
LRG2 & 0.706 &17.351 $\pm$ 0.177 & 19.455 $\pm$   0.330& 16.050 $\pm$ 0.110 \\
LRG3+ELG1 & 0.934 & 21.576 $\pm$  0.152  & 17.641  $\pm$  0.193 & 19.721 $\pm$ 0.091 \\
ELG2 & 1.321 & 27.601 $\pm$ 0.318  & 14.176  $\pm$  0.221 &24.252 $\pm$ 0.174 \\
QSO & 1.484 & 30.512  $\pm$0.760 & 12.817 $\pm$ 0.516 &26.055  $\pm$  0.398 \\
Ly$\alpha$ & 2.330 & 38.988 $\pm$ 0.531& 8.632  $\pm$ 0.101 &31.267 $\pm$ 0.256\\
\hline
\end{tabular}
\end{table}
To incorporate DESI-DR2 data into our statistical analysis, we include its contribution in the total likelihood via the chi-square function

\begin{equation}
\chi^2_{\text{DESI}} = \vec{X}_{\text{DESI}}^{\,T} \, C_{\text{DESI}}^{-1} \, \vec{X}_{\text{DESI}}, \tag{3.20}
\end{equation}
where $\vec{X}_{\text{DESI}}$ is the vector of residuals between the predicted theoretical values and the DESI observations, and $C_{\text{DESI}}$ is the corresponding covariance matrix provided by the collaboration \cite{DESI2025}.
\subsection{ PantheonPlus data}

Our cosmological analysis employs the latest PantheonPlus supernova sample, which comprises 1701 high-quality light curves from 1550 distinct type Ia supernovae \cite{PantheonPlus2022}. These observations span an extensive redshift range of $z \in  [0.001, 2.261]$, providing comprehensive coverage of cosmic expansion history. Throughout this work, we designate this compilation as the PP dataset.

The Statistical comparison between theoretical predictions and observational data is performed by minimizing the chi-square ($\chi^2$) function. For the PP dataset, the corresponding chi-square statistic is expressed as
\begin{equation}
\chi_{\mathrm{PP}}^{2} = \vec{D}^{T} \mathbf{C}_{\mathrm{PP}}^{-1} \vec{D},
\label{eq:chi2_pp}
\end{equation}
where $\mathbf{C}_{\mathrm{PP}}$ represents the full covariance matrix of the PP dataset, integrating both statistical and systematic uncertainties, and $\vec{D} = m_{Bi} - M - \mu_{\text{model}}$ \cite{PantheonPlus2022}  is the residual vector, where the index $i$ labels each individual type~Ia supernova in the PantheonPlus sample, with $m_{Bi}$ is the observed apparent magnitude (corrected for selection effects and light curve properties), $M$ is the absolute magnitude (assumed constant for standardizable candles) and $\mu_{\text{model}}$ is the model-dependent distance modulus, derived from cosmological parameters.

For a flat Friedmann Lema\^{\i}tre Robertson Walker (FLRW) Universe, the theoretical distance modulus $\mu_{\text{model}}(z)$ is computed as

\begin{equation}
\mu_{\text{model}}(z) = 5 \log_{10} \left( \frac{D_L(z)}{(H_0/c) \ \rm Mpc} \right) + 25,
\end{equation}
with the luminosity distance $D_{L}(z)$ given by

\begin{equation}
D_{L}(z) = (1 + z) \int_{0}^{z'} \frac{dz}{E(z')}.
\label{eq:luminosity_distance}
\end{equation}

A key innovation of PP is its use of Cepheid-calibrated host galaxies from the SH0ES Cepheid-host sample \cite{Riess2022} to independently constrain the absolute magnitude $M$,and break the degeneracy between $H_{0}$ and $M$. This is achieved by partitioning the residual vector $\mathbf{D}$ in Eq.~\ref{eq:chi2_pp}

\begin{equation}
\vec{\textbf{D}'_i} = 
\begin{cases} 
m_{Bi} - M - \mu^{\text{Ceph}}_i & \text{(Cepheid hosts)}, \\ 
m_{Bi} - M - \mu_{\text{model}}(z_i) & \text{(otherwise)},
\end{cases}
\label{eq:residual_vector}
\end{equation}
where $\mu^{\text{Ceph}}_i$ is the geometric distance modulus derived from Cepheid variables in host galaxies of Type Ia supernovae. 
After applying the Cepheid-based calibration implemented in the PantheonPlus analysis, the chi-square estimator is expressed in terms of the calibrated residual vector $\vec{D}'$ as
\begin{equation}
\chi^2_{\mathrm{PP}} = {\vec{D'}}^{\,T}\,
\mathbf{C}_{\mathrm{PP}}^{-1}\,
\vec{D'} .
\end{equation}
\subsection{The $H(z)$ measurements}
To further constrain the cosmological models, we incorporate direct measurements of the Hubble parameter $H(z)$. These measurements are obtained either from the clustering of galaxies and quasars (using BAO in the radial direction) \cite{Moresco2016,Alam2017} or from the differential age method, which expresses the Hubble parameter as \cite{Jimenez2002}
\begin{equation}
    H(z)=-\frac{1}{1+z} \frac{dz}{dt}.
\end{equation}
The $\chi^2$  for the $H(z)$ data is given by
\begin{equation}
    \chi^2 _{H(z)}=\sum_{i}^{36}  \left(\frac{H_{theo}(z_{i})-H_{obs}(z_{i})}{\sigma_{i}}\right)^2,
\end{equation}
where $H_{theo}(z_{i})$ is the theoretical values, $H_{obs}(z_{i})$ is the observed values at $z_{i}$ and $\sigma_{i}$ is the standard deviation.\\

We construct the total $\chi^2$ for cosmological parameter estimation by summing the individual
dataset contributions

\begin{equation}
    \chi_{\rm tot}^2 = \chi_{\rm CMB}^2 + \chi_{\rm DESI}^2 + \chi_{\rm PP}^2+ \chi_{H(z)}^2 .
\end{equation}

By minimizing $\chi_{\rm tot}^2$, we obtain the best-fit values for the cosmological parameters
of the running vacuum models.

To obtain the optimal constraints on the model parameters, we perform a Bayesian MCMC analysis
\cite{Lewis2002} using the combined dataset of CMB, DESI-DR2, PP, and $H(z)$. We consider seven free
parameters, namely $\{\Omega_{m0}, \Omega_{b}h^2, \alpha, \beta, \nu, h, M\}$, where $M$ is
marginalized through the PP likelihood and $\Omega_b h^2$ is constrained via the CMB
likelihood. The radiation density parameter $\Omega_r$ is fixed to $8 \times 10^{-5}$\cite{Planck2020}, following the Planck 2018. The
priors adopted in this work are summarized in Table~\ref{tab:priors}.

For model comparison, we employ statistical information criteria, including the Akaike Information Criterion AIC, the Deviance
Information Criterion DIC and their differences
$\Delta \mathrm{AIC}$ and $\Delta \mathrm{DIC}$ relative to the standard $\Lambda$CDM model.
The AIC is defined as \cite{Akaike1974}
\[
\mathrm{AIC} = \chi^2_{\rm min} + 2 k,
\]
where $k$ is the number of free parameters, and the relative difference is
\[
\Delta \mathrm{AIC} = \mathrm{AIC}_{\rm model} - \mathrm{AIC}_{\Lambda \rm CDM}.
\]
The DIC is defined as \cite{Spiegelhalter2002}
\[
\mathrm{DIC} = \bar{D} + p_D,
\]
where $\bar{D}$ is the mean deviance over the posterior distribution, and $p_D = \bar{D} -
D(\bar{\theta})$ represents the effective number of parameters, with $\bar{\theta}$ denoting the
posterior mean of the parameters.\\
The difference in DIC relative to the $\Lambda$CDM model is defined as
\[
\Delta \mathrm{DIC} = \mathrm{DIC}_{\rm model} - \mathrm{DIC}_{\Lambda\rm CDM}.
\]

Lower values of AIC and DIC
indicate a better balance between goodness of fit and model complexity. Accordingly, negative values 
of $\Delta \mathrm{AIC}$ or $\Delta \mathrm{DIC}$ imply that the running vacuum model provides a better 
global description of the data compared to the $\Lambda$CDM model, while positive values indicate a 
preference for $\Lambda$CDM.

 The statistical relevance of a model can be assessed through the differences in information criteria, such as $\Delta\mathrm{AIC}$ and $\Delta\mathrm{DIC}$. When $0 <\Delta\mathrm{AIC} < 2$, the competing models are essentially indistinguishable in terms of their support from the data. Values in the interval $2 \leq \Delta\mathrm{AIC} < 4$ indicate a modest preference against the model with the larger information criterion. A range of $4 \leq \Delta\mathrm{AIC} < 6$ reflects clear evidence disfavoring that model, while $6 \leq \Delta\mathrm{AIC} < 10$ points to a strong level of disfavor. If $\Delta\mathrm{AIC} \geq 10$, the data provide overwhelming evidence against the model with the higher $\mathrm{AIC}$. The same criteria apply when interpreting $\Delta\mathrm{DIC}$.

The free parameters for $\Lambda$CDM are $\{\Omega_{m0},\Omega_{b}h^2,h,M\}$, and the free parameters for the models A, B, and C are $\{\Omega_{m0}, \Omega_{b}h^2,\alpha,\beta,\nu, h, M\}$, $\{\Omega_{m0},\Omega_{b}h^2,\alpha,\nu,h, ,M\}$ and $\{\Omega_{m0},\Omega_{b}h^2,\alpha,\beta=\nu,h,M\}$, respectively.
The priors imposed on the different cosmological parameters used in our analysis are summarized in Table ~\ref{tab:priors}.
\begin{table}[H]
\centering
\caption{Priors imposed on different cosmological parameters.}
\label{tab:priors}
\begin{tabular}{l|l}
\hline
\textbf{Parameter} & \textbf{ \hspace{1cm}Prior (Model)} \\[0.1cm] 
\hline
$\Omega_{m0}$       & [0.2, 0.5]   \\ [0.1cm]
$\Omega_b h^2$   & [0.005, 0.1] \\[0.1cm]
$\alpha$         & [-0.25, 0.15] \hspace{0.3cm}(Model A) \\
                 & [0, 0.2]      \hspace{1cm}(Model B) \\
                 & [-0.2, 0.2]   \hspace{0.62cm}(Model C) \\[0.1cm]
$\beta$          & [-0.25, 0.15] \\[0.1cm]
$\nu$            & [-2, 2]        \hspace{1.2cm}(Model A) \\
                 & [0, 0.2]      \hspace{0.95cm} (Model B) \\
                     & [-0.2, 0.2]   \hspace{0.7cm}(Model C)\\[0.1cm]
$H_{0}$              & [60, 80] \\[0.1cm]
$M$              & [-20, -19] \\[0.1cm]
\hline
\end{tabular}
\end{table}
In the following, we present the results obtained from the MCMC analysis  for three cases, namely  model A, model B and model C .

\section{RESULTS AND DISCUSSIONS}
\label{sec:results}

The Markov Chain Monte Carlo analysis provides comprehensive constraints on the
cosmological parameters for the three running vacuum models under investigation,
alongside the standard $\Lambda$CDM framework. The results are derived using the combined
dataset CMB, DESI-DR2, PP, and $H(z)$, which allows us to assess both the
robustness of the models and the impact of high-precision DESI-DR2 data on the inferred
cosmological parameters.

\begin{figure}[H] 
    \centering  \includegraphics[width=1\textwidth,height=1\textheight,keepaspectratio]{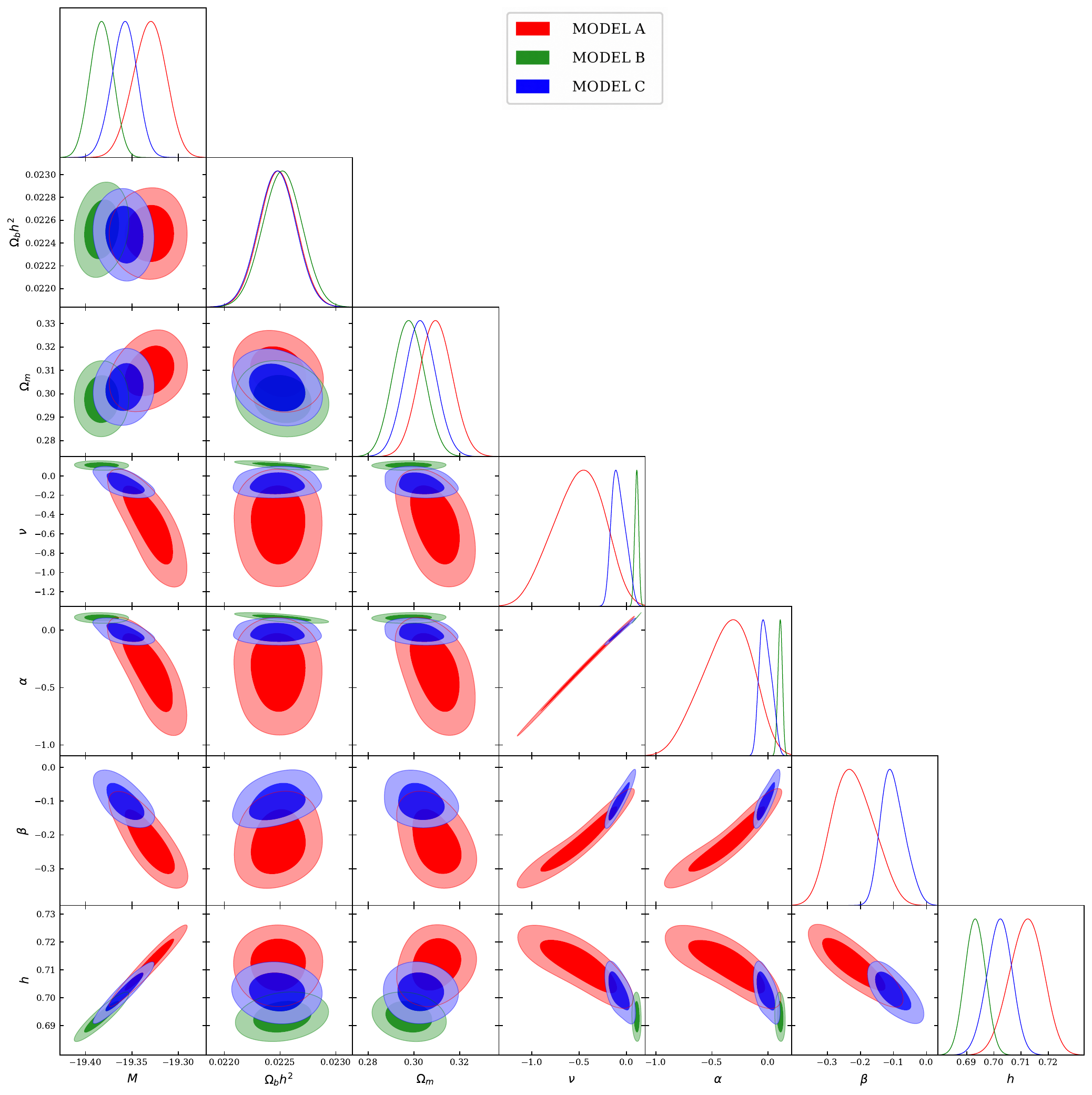}
    \caption{The $1\sigma$ and  $2\sigma$ confidence contours obtained from CMB+DESI-DR2+PP+H(z) data for the models A, B and C.} 
\label{fig:graphique1}
\end{figure}

 \begin{table*}[!htp]
\centering
\caption{Mean values $\pm 1\sigma$ of cosmological parameters, $\chi^2_{\rm min}$, $\Delta$AIC, and $\Delta$DIC for the $\Lambda\text{CDM}$ and RVMs using the CMB+DESI-DR2+PP+H(z) data combination.}
\begin{tabular}{c|c|c|c|c}
\hline
\hline
\multicolumn{1}{c|}{Data} & \multicolumn{4}{c}{CMB+DESI-DR2+PP+H(z) } \\
\hline
\multicolumn{1}{c|}{Model} & $\Lambda$CDM & Model A & Model B & Model C \\
\hline
$\Omega_{m0}$ & 0.29639 ± 0.00013 & 0.3095 ± 0.0070 & 0.2978 ± 0.0066 & 0.3027 ± 0.0066 \\[0.1cm]
$\Omega_b h^2$ & 0.02264 ± 0.0035 & 0.02248 ± 0.00016 & 0.02251 ± 0.00017 & 0.02247 ± 0.00016 \\[0.1cm]
$H_0$ [km s$^{-1}$ Mpc$^{-1}$] & 68.84 ± 0.29 & 71.19 ± 0.60 & 69.31 ± 0.36 & 70.20 ± 0.45 \\[0.1cm]
$\beta$ & - & -0.222$^{+0.057}_{-0.058}$ & 0 & -0.101$^{+0.031}_{-0.041}$  \\[0.1cm]
$\alpha$ & - & -0.37$^{+0.25}_{-0.20}$ & 0.108 ± 0.019 & -0.021$^{+0.045}_{-0.062}$  \\[0.1cm]
$\nu$ & - & -0.51 $^{+0.29}_{-0.24}$ & 0.109 ± 0.020 & - \\[0.1cm]
$M$ & -19.3978 ± 0.00088 & -19.331 ± 0.017 & -19.383 ± 0.012 & -19.358 ± 0.013 \\[0.1cm]
\hline
\multicolumn{5}{c}{ \hspace{3cm} Statistical results} \\[0.1cm]
\hline
$\chi^2_{\rm min}$ & 1592.45 & 1578.91 & 1582.34 & 1579.01 \\[0.1cm]
$\Delta$AIC & 0 & -7.535 & -6.113 & -9.440 \\[0.1cm]
$\Delta$DIC & 0 & -6.279 & -4.469 & -8.425 \\[0.1cm]
\hline
\hline
\end{tabular}
\label{tab:DESI_results}
\end{table*}
 The best-fit cosmological parameters for the three running vacuum models obtained using the combined dataset of CMB, DESI-DR2 measurements, PP, and $H(z)$,  are
summarized in Table~\ref{tab:DESI_results}. The corresponding two-dimensional confidence contours at the $1\sigma$ and $2\sigma$ levels for models A, B, and C are shown in Fig.~\ref{fig:graphique1}.

A key outcome of our analysis is that the inclusion of DESI-DR2 data leads to tight constraints on the late time expansion history, thereby significantly restricting the allowed parameter space of all models. For $\Lambda$CDM, the matter density parameter is tightly constrained around, $\Omega_{m0}= 0.29639 \pm 0.00013$, with a Hubble constant, $H_{0}=68.84 \pm0.29$ $\mathrm{km.s^{-1}.Mpc^{-1}}$. When vacuum dynamics are allowed, as in RVMs, the preferred matter density shifts to slightly higher values, namely $\Omega_{m0}=0.3095 \pm 0.0070 $ for Model~A, $\Omega_{m0}=  0.2978 \pm 0.0066 $ for Model~B, and $\Omega_{m0}= 0.3027 \pm 0.0066$ for Model~C. This behavior reflects the fact that part of the late time acceleration can be driven by dynamical evolution of the running vacuum component rather than by a strictly constant cosmological term. Importantly, the baryon density $\Omega_b h^2$ remains consistent across all models, indicating that early time physics constrained by CMB is essentially unaffected by the running vacuum parameters.

The Hubble constant inferred within the running vacuum  framework is generally higher than the value predicted by the standard $\Lambda$CDM model, with best--fit values of $H_0 = 71.19 \pm 0.60\ \mathrm{km.s^{-1}.Mpc^{-1}}$, $69.31 \pm 0.36\ \mathrm{km.s^{-1}.Mpc^{-1}}$, and $70.20 \pm 0.45\ \mathrm{km.s^{-1}.Mpc^{-1}}$ for Models~A, B, and~C, respectively. To quantify the level of agreement with the $\Lambda$CDM prediction, we evaluate the statistical deviation  estimator $T(H_0)$ \cite{Riess2019}. We obtain deviations of approximately $3.5\sigma$, $1\sigma$, and $2.5\sigma$ for Models~A, B, and~C, respectively. Therefore, Model~A exhibits a statistically significant deviation from the $\Lambda$CDM prediction, whereas Model~B remains in very good agreement with the standard cosmological scenario and Model~C shows only a moderate deviation. The relatively high values of $H_0$, particularly for Model~A, further motivate the investigation of running vacuum models as possible alternatives for alleviating the Hubble tension \cite{DiValentino2021}.

Turning to the  running vacuum parameters, we find that Model~A , which allows $\alpha$, $\beta$, and $\nu$ to vary independently, exhibits relatively broad posterior distributions, with best--fit values $\alpha = -0.37^{+0.25}_{-0.20}$, $\beta =  -0.222^{+0.057}_{-0.058} $, and $\nu =  -0.51^{+0.29}_{-0.24} $. This reflects the higher flexibility of the model, which can accommodate mild dynamical effects of the running vacuum without forcing strong deviations from the $\Lambda$CDM limit. Models~B  and~C, which impose additional relations among the  running vacuum parameters, lead to more tightly constrained results. In Model~B,  $\alpha = 0.108 \pm 0.019$ and $\nu = 0.109 \pm 0.020$ are clearly nonzero but remain small, indicating a weak dependence of the running vacuum energy on $H^2$ and $\dot{H}$. In Model~C, the parameters are constrained to $\alpha =-0.021^{+0.045}_{-0.062}$ and $\beta (=\nu) = -0.101^{+0.031}_{-0.041}$, again suggesting mild vacuum dynamics that remain compatible with current cosmological observations and previous studies of running vacuum models \cite{Sola2017,SolaPeracaula2023}.

The two-dimensional confidence contours shown in Fig.~\ref{fig:graphique1} provide additional insight into the parameter correlations. In Model~A, pronounced negative correlations are observed between the pairs $\{\alpha,\beta\}$, $\{\alpha,\nu\}$, and $\{\beta,\nu\}$, indicating the presence of parameter degeneracies. Such correlations imply that variations in one parameter can be partially compensated by opposite variations in another while yielding similar predictions for the cosmic expansion history. In contrast, the additional constraints imposed in Models~B and~C significantly reduce these degeneracies, resulting in smaller confidence regions and tighter parameter constraints.

From a statistical perspective, all three RVMs achieve lower minimum chi--square values than $\Lambda$CDM, with $\chi^2_{\rm min} = 1578.91$ (Model~A), $1582.34$ (Model~B), and $1579.01$ (Model~C) compared to $1592.45$ for $\Lambda$CDM, indicating an improved raw fit to the combined dataset. When model complexity is taken into account through AIC and DIC, all three running vacuum models show significantly negative values, namely $\Delta$AIC $= -7.535$, $-6.113$, and $-9.440$, together with $\Delta$DIC $= -6.279$, $-4.469$, and $-8.425$ for Models~A, B, and C, respectively. According to standard selection criteria, these differences reflect a clear improvement in the statistical fit provided by RVMs compared to $\Lambda$CDM . This indicates that the RVMs offer a more robust description of the current combined dataset when information criteria are applied.

Overall, these results indicate that the joint analysis of current high precision observations, allows for mild dynamical vacuum effects without strictly requiring them. From a physical perspective, this behavior can be interpreted as an effective interpolation between a purely constant running vacuum energy and a slowly evolving dark energy component. In this sense, the running vacuum framework naturally encompasses $\Lambda$CDM as a limiting case, while remaining sufficiently flexible to capture small departures from a constant cosmological term that may arise from quantum effects in curved spacetime. As a result, RVMs provide a theoretically well motivated and observationally viable extension of the standard cosmological model, fully consistent with present data and capable of accommodating future high-precision constraints.

\begin{figure}[H]
\centering
\includegraphics[width=0.65\textwidth]{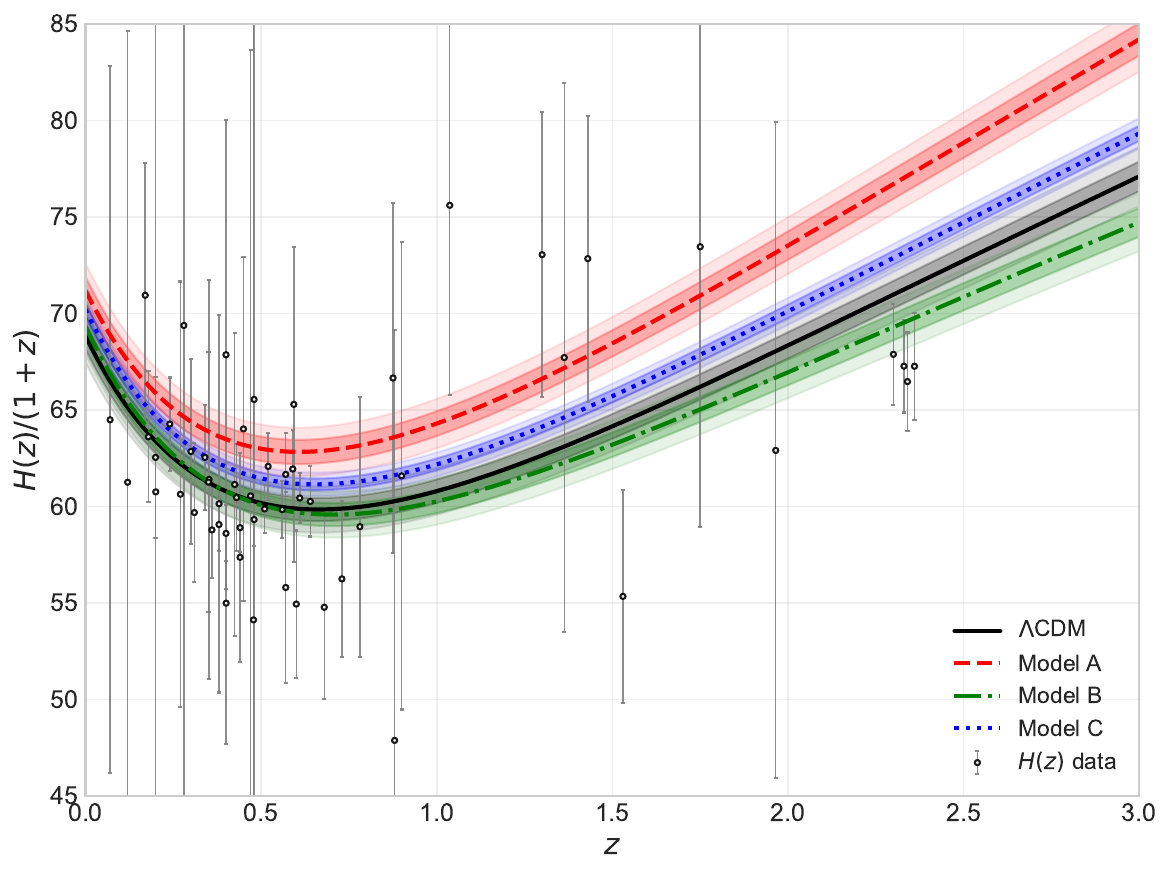}
\caption{The figure shows the evolution of the normalized Hubble parameter $H(z)/(1+z)$ as a function of redshift $z$ for $\Lambda$CDM and RVMs, together with their corresponding $1\sigma$ and $2\sigma$ confidence intervals. Points with error bars correspond to the observational $H(z)$ measurements \cite{Farooq2017}.}
\label{Fig1_Hz.pdf}
\end{figure}

\begin{figure}[H]
\centering
\begin{minipage}[b]{0.48\textwidth}
\includegraphics[height=7cm]{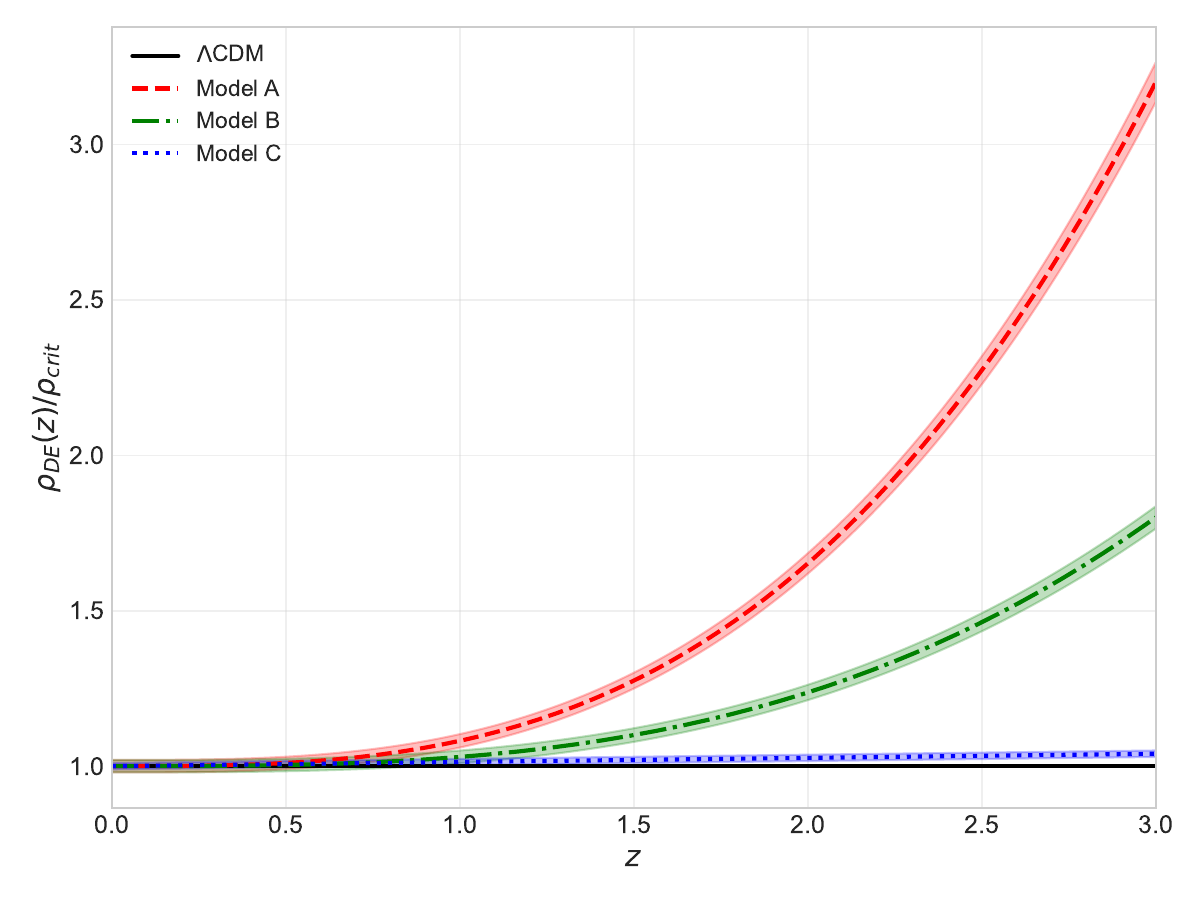} 
\caption{The figure shows the normalized dark energy density $\rho_{\rm DE}(z)/\rho_{\rm crit}$ as a function of redshift $z$ for the $\Lambda$CDM model and RVMs, together with their $1\sigma$ and $2\sigma$ confidence intervals.}
\label{Fig2_rhoDE.pdf}
\end{minipage}
\hfill
\begin{minipage}[b]{0.48\textwidth}
\includegraphics[height=7cm]{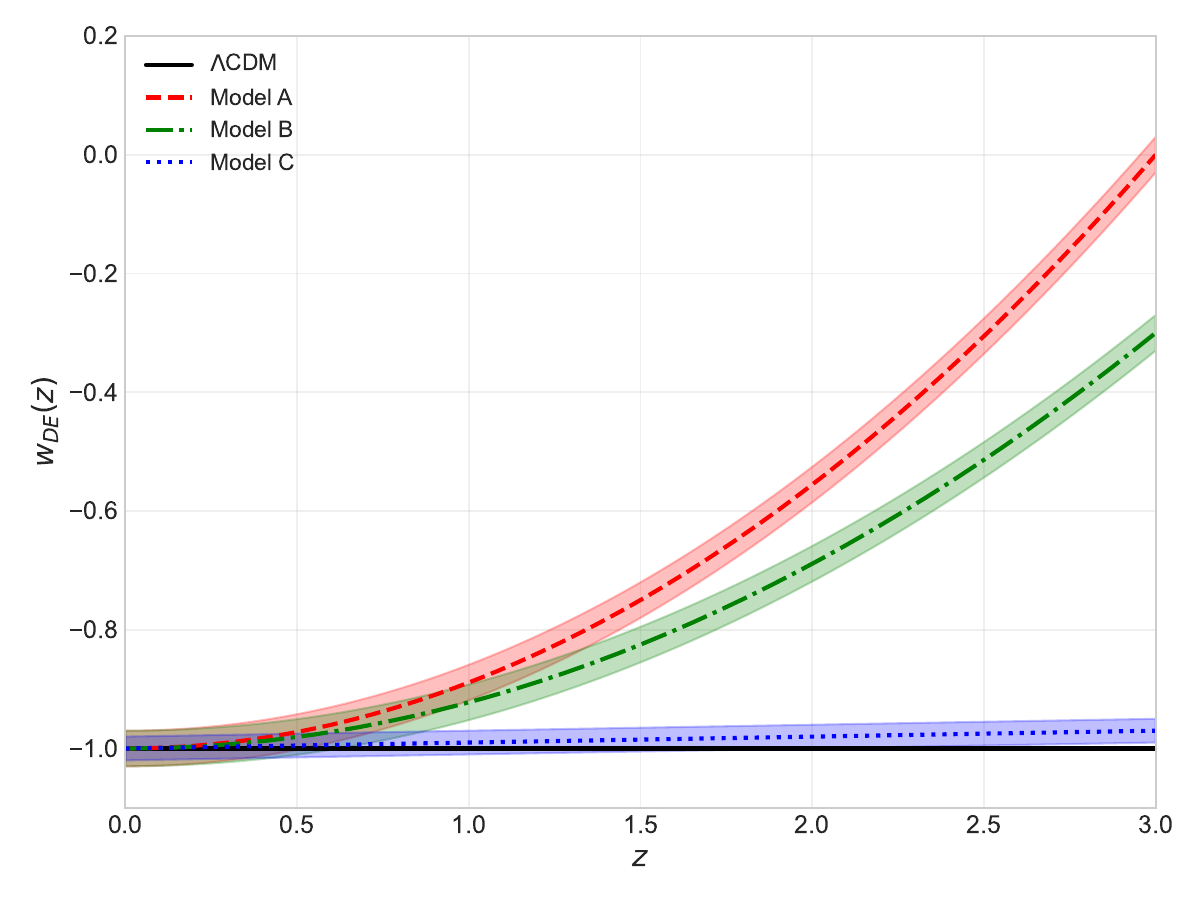}
\caption{The figure shows the effective EoS parameter $w_{\rm DE}(z)$ as a function of redshift $z$ for the $\Lambda$CDM model and RVMs, together with their $1\sigma$ and $2\sigma$ confidence intervals.}
\label{fig3_wDE.pdf}
\end{minipage}
\end{figure}

To illustrate the behavior of the running vacuum models compared to $\Lambda$CDM, we plot the normalized Hubble parameter $H(z)/(1+z)$, the dark energy density $\rho_{\rm DE}/\rho_{\rm crit}$, and the effective equation of state parameter, $w_{\rm DE}$, as functions of redshift $z$. These plots highlight the evolution of the running vacuum energy and its influence on the cosmic expansion. As shown in Fig.~\ref{Fig1_Hz.pdf}, \ref{Fig2_rhoDE.pdf}, and \ref{fig3_wDE.pdf}

The evolution of the normalized Hubble parameter $H(z)/(1+z)$, shown in Fig.~\ref{Fig1_Hz.pdf}, indicates that all running vacuum models closely follow the $\Lambda$CDM prediction at low redshifts ($z \lesssim 0.5$), where observational constraints are strongest. At intermediate and higher redshifts, deviations from the standard cosmological scenario become more pronounced, reflecting the impact of vacuum dynamics on the expansion history. Model~A exhibits the largest deviation from $\Lambda$CDM, corresponding to a higher expansion rate at increasing redshift. In contrast, Model~B remains the closest to the $\Lambda$CDM prediction over the full redshift range, while Model~C exhibits an intermediate behavior between Models~A and~B. When compared with the observational $H(z)$ measurements, all models remain compatible with current data within the observational uncertainties.


Fig.~\ref{Fig2_rhoDE.pdf} presents the redshift evolution of the normalized dark energy density, $\rho_{\rm DE}(z)/\rho_{\rm crit}$. The $\Lambda$CDM model predicts a constant vacuum energy density, whereas the running vacuum models allow for a small redshift dependence. Model~A exhibits the strongest deviation from the standard cosmological scenario, leading to a rapid increase of the dark energy density at higher redshifts. Model~B also shows a noticeable but more moderate evolution, whereas Model~C remains very close to a constant vacuum energy density and closely tracks the $\Lambda$CDM behavior over the full redshift range. These results illustrate that the running vacuum framework naturally interpolates between a strictly constant vacuum energy and a slowly evolving dark energy component.

The effective equation of state parameter $w_{\rm DE}(z)$, shown in Fig.~\ref{fig3_wDE.pdf}, provides further insight into the dynamical properties of the running vacuum models. At low redshifts, all models converge to $w_{\rm DE}\simeq -1$, consistent with the cosmological constant. As the redshift increases, Models~A and~B gradually evolve toward $w_{\rm DE}>-1$, exhibiting a quintessence-like behavior, with a more pronounced deviation in Model~A. In contrast, Model~C remains very close to $w_{\rm DE}=-1$ over the entire redshift range, making its evolution nearly indistinguishable from that of the $\Lambda$CDM model. 

\section{Conclusion}
\label{sec:Conclusion}
In this work, we have investigated an extended class of running vacuum models in which the vacuum energy density evolves dynamically with the Hubble parameter and its time derivatives, providing a natural generalization of the standard cosmological constant paradigm. By introducing a dynamical vacuum equation of state and deriving the corresponding modified Friedmann equations, we obtained a consistent framework that naturally reduces to $\Lambda$CDM in the limit of vanishing running parameters.

We confronted three different realizations of the running vacuum scenario with the latest cosmological observations, combining CMB shift parameters, DESI-DR2 dataset, PantheonPlus supernovae, and $H(z)$ measurements. A full Bayesian MCMC analysis was performed to constrain the cosmological and running vacuum parameters to assess the statistical performance of each model relative to $\Lambda$CDM.

Our results demonstrate that all three RVMs provide an excellent fit to current cosmological data and remain fully consistent with the standard $\Lambda$CDM scenario. In particular, the matter and baryon density parameters, as well as the Hubble constant, are compatible with $\Lambda$CDM values, while the vacuum parameters exhibit small but non-negligible deviations, reflecting the potential influence of dynamical vacuum effects. Model~A, which allows all running vacuum parameters to vary independently, exhibits the strongest deviations from the $\Lambda$CDM limit, whereas Model~B shows only mild deviations, and Model~C displays the smallest departures from the standard cosmological scenario.

The running vacuum framework naturally interpolates between a constant vacuum energy and a slowly evolving dark energy component, providing a theoretically motivated and observationally viable extension of the standard cosmological model. The combined cosmological dataset considered in this work, including the latest DESI-DR2 measurements, provides tight constraints on the running vacuum parameters and allows a robust assessment of their cosmological viability.

From a statistical perspective, all three running vacuum models achieve lower minimum chi-square values than the standard $\Lambda$CDM model. Furthermore, the information criteria indicate a mild statistical preference for the running vacuum scenarios relative to $\Lambda$CDM. However, further high precision observations will be required to establish the robustness of this preference and to determine whether the observed deviations are signatures of genuine vacuum dynamics.

In summary, these findings indicate that RVMs constitute a flexible and well motivated alternative to $\Lambda$CDM, capable of accommodating mild vacuum dynamics while preserving excellent agreement with current high precision observational data. These results provide a direct observational test of RVMs and show that they remain viable and compatible with current cosmological observations. Future high precision surveys will allow further exploration of potential departures from a strictly constant running vacuum energy, offering new insights into the nature of dark energy and the underlying physics of the Universe.


\end{document}